# NEAR STELLAR SOURCES OF GAMMA-RAY BURSTS

B.I.Luchkov, P.D.Markin
National Research Nuclear University MEPhI
Moscow, Kashirskay Road 31

Correlation analysis of gamma-ray burst coordinates and nearby stars, registered on 2008-2011, revealed 5 coincidences with angular accuracy better than $0.1^o$. The random probability is $7\times10^{-7}$, so evidencing that coincident stars are indeed gamma-ray burst sources. The proposed method should be continued in order to provide their share in common balance of cosmic gamma-ray bursts.

Cosmic gamma-ray bursts (GRB), short gamma-ray fluxes with 0.01-10 MeV energy interval, registered by space devices, are investigated more than 30 years [1, 2]. Their sources and distances were not known for a long time. Some terrestrial telescopes as well as Hubble space telescope obtained fading objects with great red shifts ($Z \geq 1$) [3]. It is evidence that GRB came from large (cosmological) distances, almost equal to Universe visual horizon. However cosmological sources, appearing as Super Nova flares in remote galaxies [4], contain only 40% of all GRB. What are the others? We discussed possibility for nearby stellar sources [5] – active stars of G, K, M spectral class known as flaring low mass stars [6]. This work is devoted to continuation of stellar gamma-ray burst research.

**Correlation analysis of gamma-ray burst and star locations**

Gamma-burst positions, registered during 2008 – 2011 by Swift, Integral, Fermi and Agile devices, are compared with stars near Sun (catalog Gliese [7]). In order to minimize random coincident amount, which sharply rise for large r, only stars with parallaxes P > 0.04 (corresponding to r ≤ 25 pc) are taken for analysis. The stars of G, K, M spectral classes should be GRB sources due to their activity. The cosmological one that may coincide with stars only randomly serves as a comparison criterion.

Numerical indicator for gamma-ray burst and star coordinates coincidence is a deviation $\Delta r = (\Delta\alpha^2 + \sin\delta\,\Delta\delta^2)^{1/2}$, where $\Delta\alpha$ and $\Delta\delta$ are angular differences for right ascension and declination. The analysis result is given in table, where small $\Delta r$ are shown.

| $\Delta r$ | $0^o$-$0.1^o$ | $0.1^o$-$0.2^o$ | $0.2^o$-$0.3^o$ | $0.3^o$-$0.4^o$ | $0.4^o$-$0.5^o$ | $0.5^o$-$0.6^o$ | $0.6^o$-$0.7^o$ | $0.7^o$-$0.8^o$ |
|---|---|---|---|---|---|---|---|---|
| Stellar GRB | 2 | 3 | 0 | 3 | 2 | 1 | 2 | 5 |
| Cosmological GRB (Z≥1) | 0 | 0 | 1 | 0 | 1 | 0 | 0 | 5 |

As one can see, 5 stellar GRBs were found: two with $\Delta r = 0 – 0.1^o$ and three with $\Delta r = 0.1 – 0.2^o$. These intervals include no cosmological gamma-ray burst! Beginning with $\Delta r = 0.2^o$ cosmological bursts, considered as comparison, appear what indicates to random superposition of background increasing at greater $\Delta r$.

**Result analysis**

Here is the list of exact gamma-ray burst and star coincidences: $\Delta r = 0.1^o$, $0.13^o$, $0.03^o$, $0.14^o$, $0.08^o$. Mean value $\Delta r = 0.1^o$ what well corresponds to devices angular accuracy. There are no similar small coincidences in control cosmological group.

Strong gamma-ray burst on May 25 2008 year of star EV Lacarta registered by tens detectors [8] became very important evidence for stellar origin. That long expected from indirect data event visually confirms near star participation in gamma-bursts. EV Lacarta is well-known nearby (r = 5 pc) flaring star of M3.5 spectral classis [6]. Coincidence accuracy was $\Delta r = 0.08^o$.

Let us give variety estimation value for 5 found coincidences. The number of random coincident events is $N_{ren} = \sigma_{coin} N_b N_{st} / \Omega = 0.16$, where $\sigma_{coin} = \pi \Delta r^2 = 0.0314$ degree$^2$. Burst and star numbers are respectively $N_b = 450$, $N_{st} = 470$. Total sky surface coverage is $\Omega = 41253$ degree$^2$. Poisson random probability for N = 5 burst-star coincidences is equal to $W = e^{-N_{ren}} N_{ren}^N / N! = 7 \times 10^{-7}$, extremely small value.

The made search for stellar gamma-ray bursts obviously showed that additional source alternative to cosmological (Z ≥ 1) is stars in nearby solar environment. The obtained coincident number is so far small, but could be, at least, all 60% of GRB still unidentified. Therefore it is necessary to continue stellar gamma-burst search both by recent method and by new better one which should allow star investigation at greater distances (r > 25 pc).

REFERENCES
1 O.F.Prilutskiy, I.L.Rosental, V.V.Usov  Uspekhi Phys.Nauk V.116, P.517 (1975)
2 B.I.Luchkov, I.G.Mitrofanov, I.L.Rosental  ibid. V.166, P.743 (1996)
3 K.A.Postnov  ibid  V.169, P.545 (1997)
4 J van Paradijs, P.J.Groot et al. Nature 386, 686 (1997)
5 B.I.Luchkov  Proceeding of Science Conference MEPhI-2009; arxiv.org/abs/1104.3351
6 R.E.Gershberg  Flaring stars of small masses, M.Nauka, 1985
7 W.Gliese, H.Jahress Catalogue of Nearby Stars (ARICNS database)
8 www.solstation.com/stars/ev-lac.htm